\documentclass[apjl]{emulateapj}

\slugcomment{\ }

\usepackage{color}
\usepackage{tabularx}
\definecolor{gg}{rgb}{0,0.6,0}
\definecolor{rr}{rgb}{0.6,0,0}

\def\adp{$\delta_{\rm ADP,N}$}

\defcitealias{dario2014}{DTJ14}

\shorttitle{The structural evolution of forming and early stage star clusters}
\shortauthors{Jaehnig et al. 2014}

\begin{document}
\title{The structural evolution of forming and early stage star clusters}
\author{Karl O. Jaehnig$^1$, Nicola Da Rio$^1$, Jonathan C. Tan$^{1,2}$}
\affil{$^1$Department of Astronomy, University of Florida, Gainesville, FL 32611, USA.\\
$^2$Department of Physics, University of Florida, Gainesville, FL 32611, USA.}
\email{k.jaehnig@yahoo.com,ndario@ufl.edu,jt@astro.ufl.edu}

\begin{abstract}
We study the degree of angular substructure in the stellar position
distribution of young members of Galactic star-forming regions,
looking for correlations with distance from cluster center, surface
number density of stars, and local dynamical age. To this end
we adopt the catalog of members in 18 young ($\sim$1--3~Myr) clusters
from the \emph{Massive Young Star-Forming Complex Study in Infrared
  and X-ray} (MYStIX) Survey and the statistical analysis of the
Angular Dispersion Parameter, \adp.  We find statistically significant
correlation between \adp\ and physical projected distance from the
center of the clusters, with the centers appearing smoother than the
outskirts, consistent with more rapid dynamical processing on local
dynamical, free-fall or orbital timescales.  Similarly, smoother
distributions are seen in regions of higher surface density,
  or older dynamical ages. These results indicate that dynamical
processing that erases substructure is already well-advanced in young,
sometimes still-forming, clusters. Such observations of the
dissipation of substructure have the potential to constrain
theoretical models of the dynamical evolution of young and forming
clusters.
\end{abstract}



\section{Introduction}
\label{section:introduction}

It is well known that the majority of stars form in aggregates or
clusters \citep{ladalada2003,gutermuth2009}, which often do not
survive as bound systems after early dynamical evolution and/or gas
removal. As star formation progresses from early, more filamentary
structures \citep[e.g.,][]{andre2013} to young clusters or field
stars, understanding the evolution of the spatial and kinematic
structures of star-forming regions can provide fundamental clues on
the boundary conditions of the star formation process.

Numerical simulations suggest that virialized star clusters can form
before gas expulsion \citep[e.g.,][]{fellhauer2009}, depending on the
initial conditions and time of gas removal. Subvirial initial
conditions for stellar motions are also a possibility, as suggested by
studies of dense gas cores \citep[e.g.,][]{kirk2007}, with subsequent
dynamical evolution investigated by a number of works
\citep[e.g.,][]{scally2002,proszkow2009,allison2009,maschberger2010,parker2012,kruijssen2012}.

Different studies have adopted several metrics to measure substructure
\citep[e.g.,][]{goodwin2004,cartwright2004,gutermuth2005,schmeja2006},
e.g., the minimum spanning tree ${\cal Q}$ parameter, mass surface
density, stellar separation, and indicators of mass
segregation. \citet{parker2014a,parker2014b} showed that the use of
multiple indicators at once helps disentangle effects of dynamical
evolution from effects related to the choice of initial conditions.

In \citet{dario2014} (hereafter \citetalias{dario2014}) we presented
an analysis of the spatial distribution (as well as dynamical status)
of sources in the Orion Nebula Cluster (ONC). We introduced a metric
to measure the degree of angular subclustering in centrally concentrated
clusters: \emph{the angular dispersion parameter} (\adp),
building on previous work by \citet{gutermuth2005}.
This quantity represents the dispersion of stellar counts measured in
$N$ equal sectors from the cluster center, normalized to the standard
deviation expected from Poisson statistics. Thus, \adp$\simeq1$ for an
azimuthally random distribution, and increases in the presence of
substructure. In contrast to other popular indicators of substructure,
e.g., the minimum spanning tree ${\cal Q}$ parameter
\citep{cartwright2004}, \adp\ is sensitive to minute deviations from
isotropy within a clearly defined stellar grouping, rather than only
distinguishing among centrally concentrated versus substructured
populations.  In the ONC we detect a clear radial dependence of \adp,
with the core being systematically smoother than the outskirts, which
we interpret as result of dynamical processing.

In this paper we extend this analysis to a sample of 18 clusters,
including the ONC, from the MYStIX Survey \citep{feigelson2013}.
\S\ref{section:data} is an overview of the data used for the
analysis. Section \S\ref{section:analysis} reviews our methods to
calculate the center and the \adp\ radial variation for each
cluster. We outline our findings in \S\ref{section:results}.

\section{The data}
\label{section:data}

\begin{figure*}
\epsscale{1.12}
\plotone{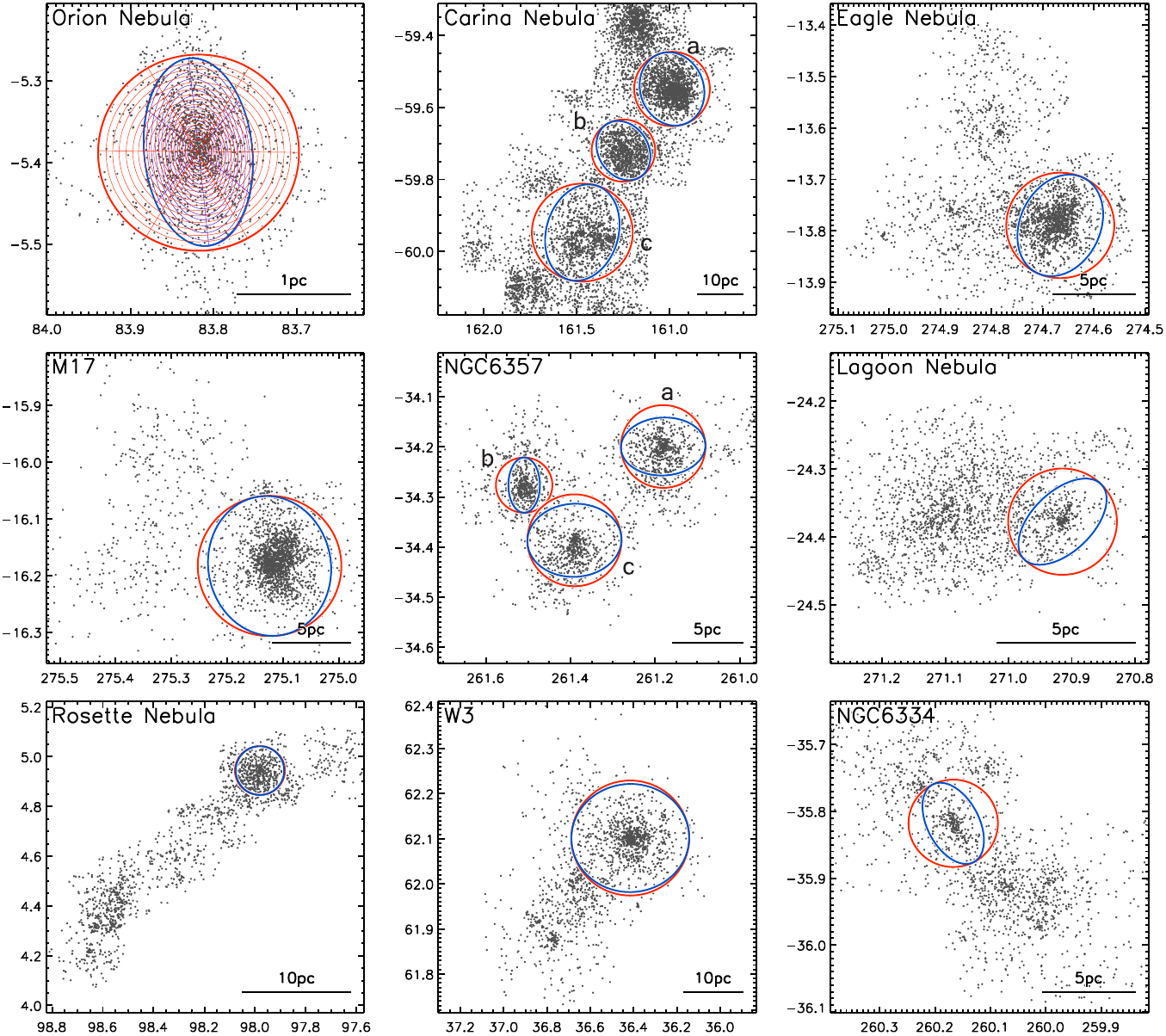}
\caption{
Spatial distribution of young candidate members in the considered
MYStIX star-forming regions. The red circles indicate the outer
cluster radius for each identified subcluster; the blue ellipses
represent the best-fit 2D Plummer ellipsoid adopted to remove the
contribution from the elongation of these structures in the
\adp\ analysis. The physical distances indicated are computed adopting
the cluster distances from \citet{feigelson2013}. In the first panel
we illustrate the subdivision of the Orion population in concentric
annuli containing 60 stars each (20 annuli for circular symmetry
(red), 17 for elliptical symmetry (blue).
-- \emph{Continues on Figure
  \ref{figure:radec2}.} \label{figure:radec1}}
\end{figure*}

\begin{figure*}
\epsscale{1.12}
\plotone{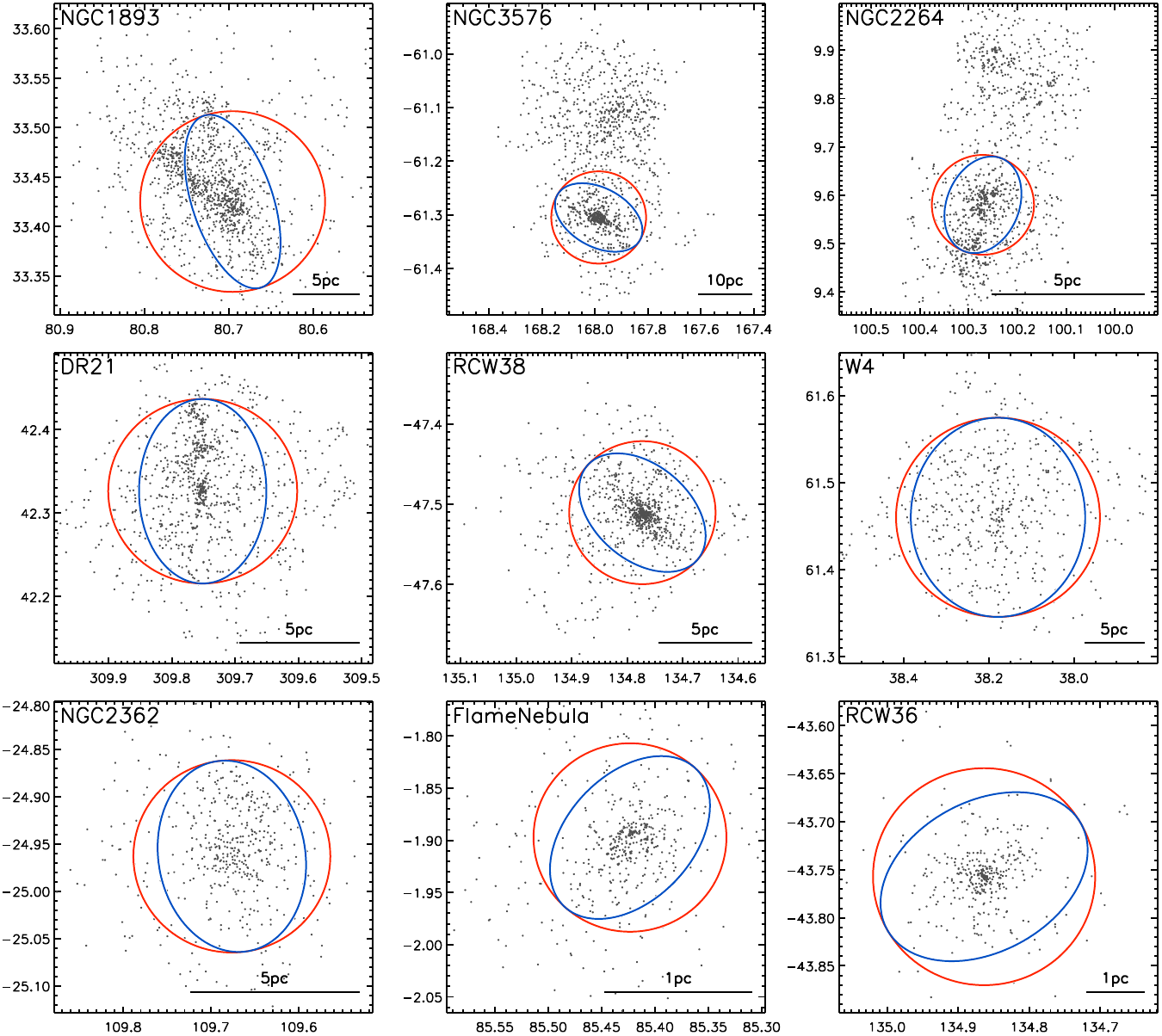}
\caption{Continued from Figure \ref{figure:radec1}. \label{figure:radec2}}
\end{figure*}

We adopt the dataset from the MYStIX survey
\citep{feigelson2013}. This survey collected data from the
\emph{Chandra} X-ray Observatory, and infrared photometry from the \emph{United Kingdom InfraRed Telescope} (UKIRT) and the
\emph{Spitzer Space Telescope} on 20 Galactic star-forming
regions with typical ages of 1--3~Myr \citep{getman2014b}. From all the these sources, \citet{broos2013} published
statistical memberships based on multiple observational
information: X-ray classification and infrared
excess from circumstellar material. We thus restricted
our analysis to these sources, for a total of 31,784 probable PMS stars. Due
to sensitivity limits and the restrictive criteria of membership assignment, the MYStIX sample is incomplete especially for low-mass stars. However, we assume that the sample is representative of the
spatial distribution of sources in each region. We exclude from our analysis two MYStIX clusters: the Trifid Nebula and W4, which
respectively are characterized by low number statistics, or lack a
definite central concentration, two aspects which undermine the
diagnostic power of \adp.

Some of the regions enclose multiple separated star clusters;
\citet{kuhn2014} identify MYStIX subclusters using collections of isothermal ellipsoids. We do not
use their identification of subclusters since some of these
groupings are poorly populated, unsuitable for \adp\ analysis, and in
many cases they indicate a stratification or core-halo distribution which could simply indicate that these systems are not well fit by a single isothermal ellipsoid. Instead, we look for clusters using a nearest
neighbor algorithm identifying overdensities at a scale of
0.065~pc.
We isolate 22 subclusters in 18 star-forming
regions, as shown in Figures \ref{figure:radec1} and
\ref{figure:radec2}. For each cluster, we defined an outer boundary
radius to be fully contained inside each Chandra field of view, and decreased it in some cases to avoid including multiple
subclusters within the same aperture.

\section{Analysis}
\label{section:analysis}

\begin{figure*}
\epsscale{1.15}
\plotone{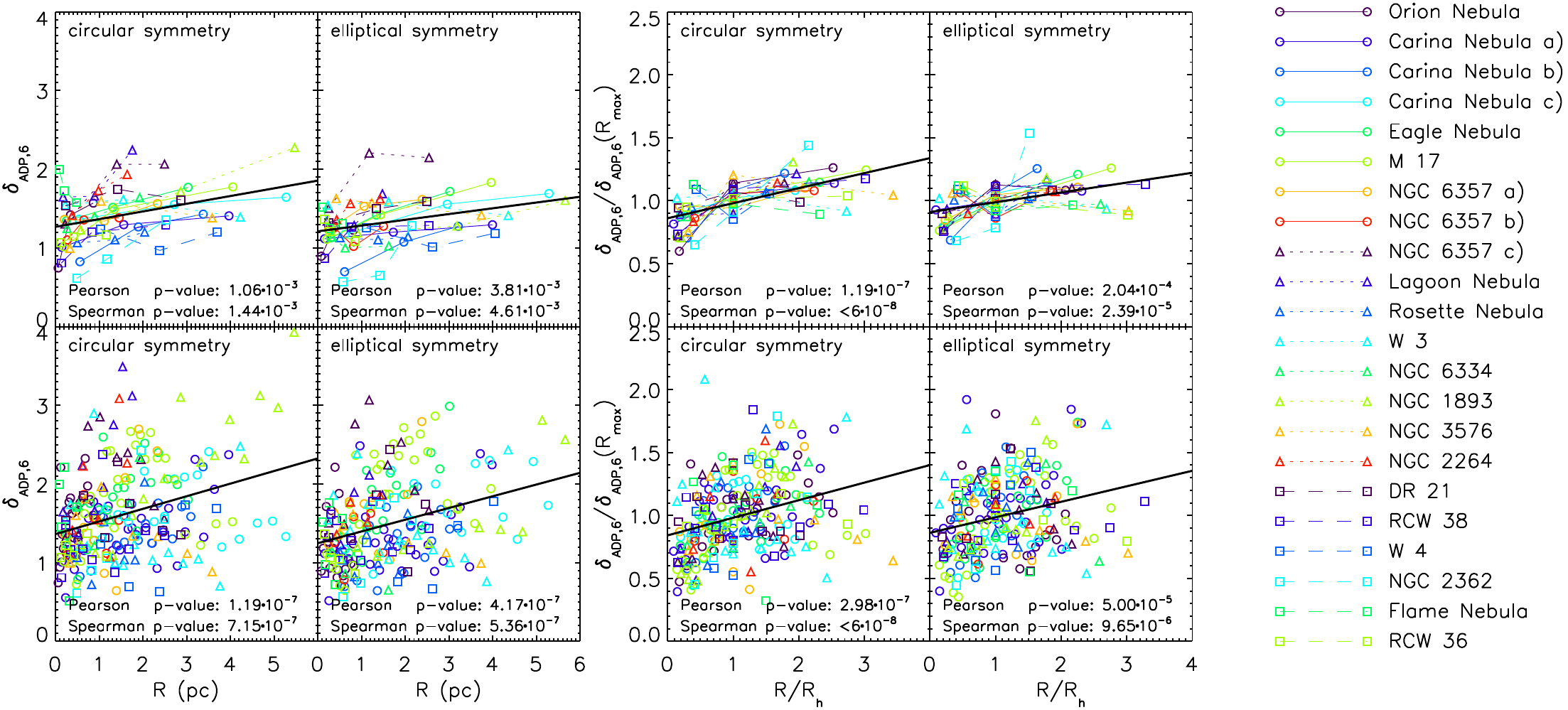}
\caption{
$\delta_{\rm ADP}$ vs $R$
for all MYStIX regions of our analysis. Top panels indicate our
low-resolution profiles (the average within 3 bins per cluster:
$R_{60}$, $R_h$ and $R_{\rm max}$), bottom panels indicate the
high-resolution profiles, where for each cluster all the radial bins
from the center (each containing 60 sources) are denoted. The
right-hand panels show the correlations between the quantities in each
axis is normalized as indicated. The thick black line is the best fit
to the data, and the correlation $p$-value is reported in each
panel. \label{figure:adp_vs_radii}}
\end{figure*}
\begin{figure*}
\epsscale{1.15}
\plotone{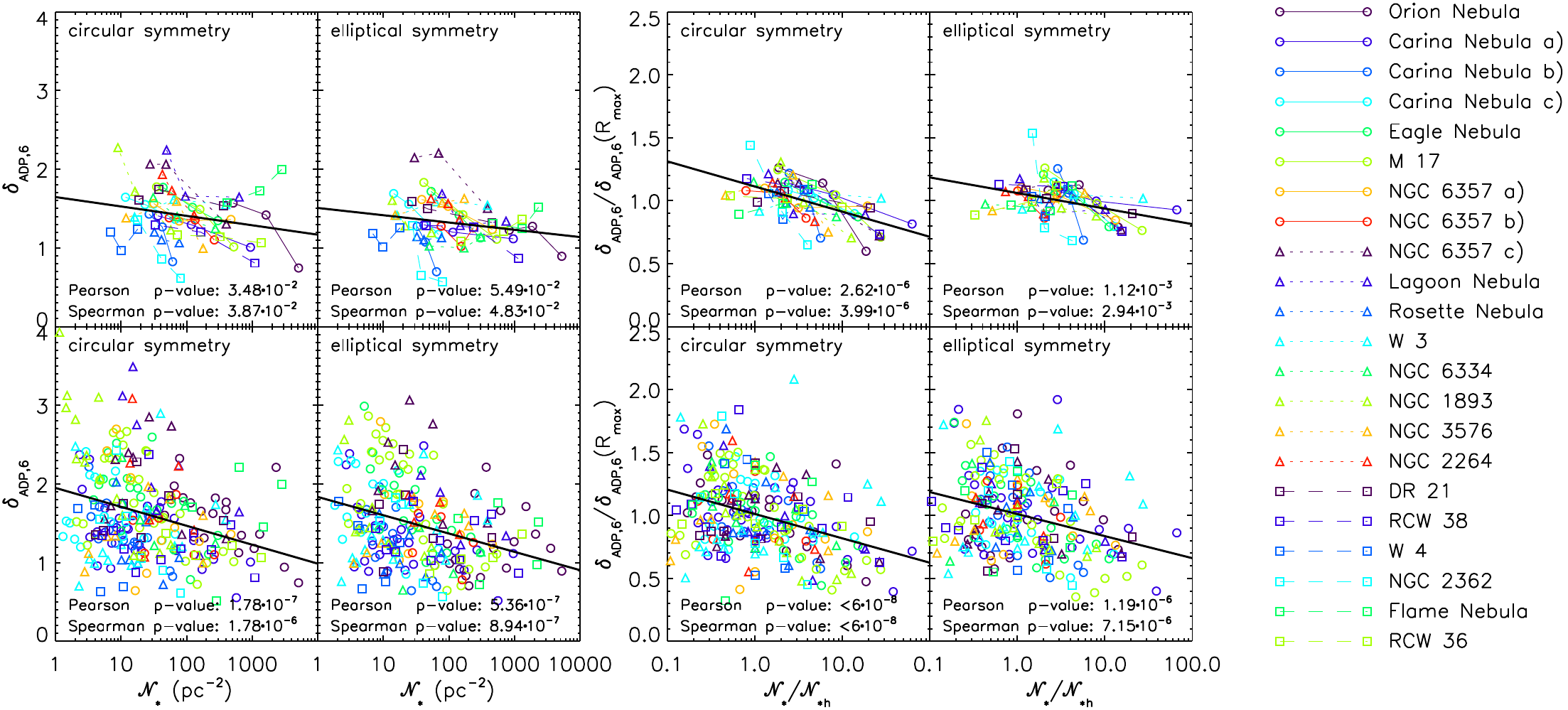}
\caption{Same as Figure \ref{figure:adp_vs_radii}, for the correlation of $\delta_{\rm ADP,6}$ with surface density of stars ${\cal N}_*$. \label{figure:adp_vs_density}}
\end{figure*}
\begin{figure*}
\epsscale{1.15}
\plotone{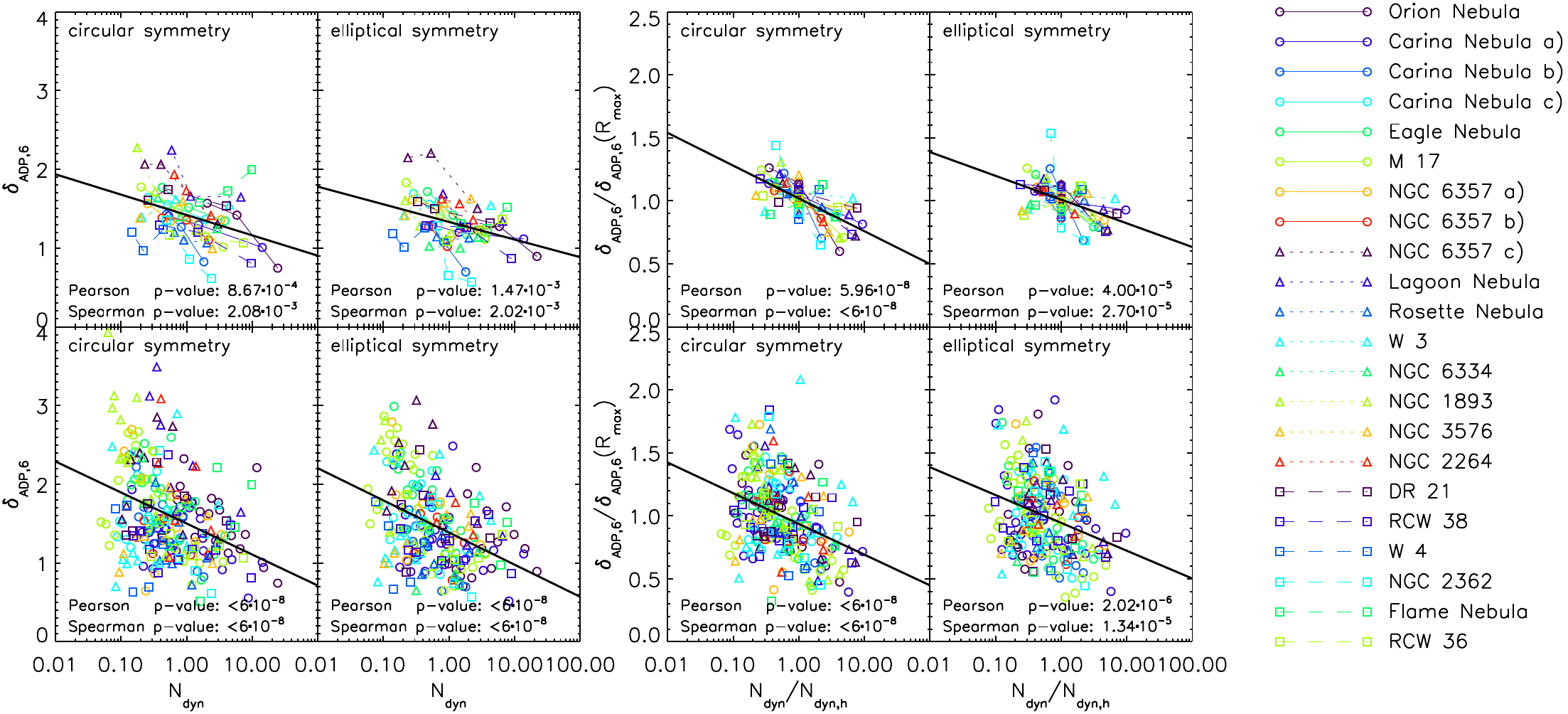}
\caption{Same as Figure \ref{figure:adp_vs_density}, for the correlation of $\delta_{\rm ADP,6}$ with dynamical age. \label{figure:adp_vs_dynage}}
\end{figure*}

\begin{table*}
\centering
\caption{ }
\label{table}
\begin{tabular}{|r|ccccccc|ccc|cc|}
\hline
\multicolumn{1}{|r}{}\vspace{-0.1cm}                    & \multicolumn{1}{|c}{$d$}  & \multicolumn{1}{c}{$t_*$}  & \multicolumn{1}{c}{$R_{60}$}     & \multicolumn{1}{c}{$R_{h}$}     & \multicolumn{1}{c}{$R_{\rm max}$} & \multicolumn{1}{c}{}              & \multicolumn{1}{c}{}                  & \multicolumn{3}{|c}{$\langle\delta_{\rm ADP,6}\rangle $}                                           & \multicolumn{2}{|c|}{MST ${\cal Q}$}                                  \\
\multicolumn{1}{|r}{Cluster} \vspace{-0.1cm}            & \multicolumn{1}{|c}{}     & \multicolumn{1}{c}{}       & \multicolumn{1}{c}{}             & \multicolumn{1}{c}{}            & \multicolumn{1}{c}{}              & \multicolumn{1}{c}{$n_{\rm tot}$} & \multicolumn{1}{c}{$N_{\rm a}$}       & \multicolumn{3}{|c}{}                                                                              & \multicolumn{2}{|c|}{}                                                \\
\multicolumn{1}{|r}{}                                   & \multicolumn{1}{|c}{(pc)} & \multicolumn{1}{c}{(Myr)}  & \multicolumn{1}{c}{(pc)}         & \multicolumn{1}{c}{(pc)}        &  \multicolumn{1}{c}{(pc)}         & \multicolumn{1}{c}{}              & \multicolumn{1}{c}{}                  & \multicolumn{1}{|c}{$R_{60}$}  & \multicolumn{1}{c}{$R_{h}$} & \multicolumn{1}{c}{$R_{\rm max}$}   & \multicolumn{1}{|c}{$R_{h}$} & \multicolumn{1}{c|}{$R_{\rm max}$}     \\[0.5ex]
\hline
Orion Nebula\vspace{-0.0cm}      & 414 & 1.6 & 0.06 & 0.34 & 0.86 & 1200 & 20 &        0.74 &     1.42 &     1.57 & 0.83 & 0.89  \\
Carina Nebula a)                & 2300 & 3.3 & 0.14 & 1.56 & 3.97 & 1800 & 30 &        1.01 &     1.30 &     1.41 & 0.79 & 0.90  \\
Carina Nebula b)                & 2300 & 3.3 & 0.56 & 1.89 & 3.38 & 960 & 16 &         0.82 &     1.27 &     1.43 & 0.76 & 0.82  \\
Carina Nebula c)                & 2300 & 3.3 & 0.54 & 2.72 & 5.27 & 1020 & 17 &        1.36 &     1.55 &     1.64 & 0.74 & 0.83  \\
Eagle Nebula                    & 1750 & 2.1 & 0.25 & 1.31 & 3.03 & 1260 & 21 &        1.35 &     1.52 &     1.77 & 0.79 & 0.90  \\
M 17                            & 2000 & 1.3 & 0.19 & 1.34 & 4.06 & 1800 & 30 &        1.01 &     1.49 &     1.78 & 0.81 & 0.99  \\
NGC 6357 a)                     & 1700 & 1.3 & 0.20 & 1.11 & 2.33 & 540 & 9 &          1.36 &     1.34 &     1.56 & 0.83 & 0.88  \\
NGC 6357 b)                     & 1700 & 1.3 & 0.27 & 0.65 & 1.45 & 360 & 6 &          1.10 &     1.36 &     1.38 & 0.80 & 0.89  \\
NGC 6357 c)                     & 1700 & 1.3 & 0.22 & 1.40 & 2.49 & 540 & 9 &          1.30 &     2.06 &     2.06 & 0.89 & 0.84  \\
Lagoon Nebula                   & 1300 & 2.1 & 0.17 & 0.89 & 1.76 & 480 & 8 &          1.65 &     1.66 &     2.24 & 0.86 & 0.84  \\
Rosette Nebula                  & 1330 & 3.1 & 0.50 & 1.36 & 2.04 & 420 & 7 &          1.07 &     1.10 &     1.20 & 0.80 & 0.80  \\
W 3                             & 2040 & ... & 0.22 & 1.55 & 4.23 & 900 & 15 &         1.55 &     1.61 &     1.39 & 0.86 & 0.92  \\
NGC 6334                        & 1700 & 1.7 & 0.26 & 0.89 & 1.92 & 360 & 6 &          1.25 &     1.40 &     1.64 & 0.82 & 0.76  \\
NGC 1893                        & 3600 & 2.6 & 0.58 & 2.86 & 5.47 & 840 & 14 &         1.23 &     1.72 &     2.27 & 0.81 & 0.79  \\
NGC 3576                        & 2800 & ... & 0.33 & 1.04 & 3.58 & 480 & 8 &          1.00 &     1.59 &     1.38 & 0.82 & 0.86  \\
NGC 2264                        &  913 & 2.4 & 0.38 & 0.98 & 1.64 & 360 & 6 &          1.41 &     1.73 &     1.93 & 0.78 & 0.76  \\
DR 21                           & 1500 & 1.9 & 0.23 & 1.43 & 2.87 & 480 & 8 &          1.53 &     1.74 &     1.61 & 0.80 & 0.80  \\
RCW 38                          & 1700 & ... & 0.13 & 0.84 & 2.52 & 660 & 11 &         0.81 &     1.20 &     1.29 & 0.89 & 0.95  \\
W 4                             & 2040 & ... & 1.03 & 2.39 & 3.70 & 300 & 5 &          1.24 &     0.97 &     1.20 & 0.81 & 0.85  \\
NGC 2362                        & 1480 & 3.5 & 0.49 & 1.18 & 2.53 & 360 & 6 &          0.61 &     0.86 &     1.36 & 0.81 & 0.87  \\
Flame Nebula                     & 414 & 1.0 & 0.08 & 0.21 & 0.48 & 300 & 5 &          2.00 &     1.73 &     1.58 & 0.74 & 0.93  \\
RCW 36                           & 700 & 1.3 & 0.12 & 0.42 & 1.16 & 300 & 5 &          1.06 &     1.14 &     1.16 & 0.88 & 0.99  \\
\hline
\end{tabular}

\begin{tabular}{|r|rrr|rrr|rrr|rrr|}
 \multicolumn{13}{r}{} \\[2ex]
\hline
\multicolumn{1}{|r}{}\vspace{-0.1cm}                    & \multicolumn{9}{|c|}{$\delta_{\rm ADP}^\prime=$   }                                                                                                                                                                                                                       & \multicolumn{3}{c|}{}           \\
\multicolumn{1}{|r}{}\vspace{-0.05cm}                   & \multicolumn{9}{|c|}{                            }                                                                                                                                                                                                                        & \multicolumn{3}{c|}{}           \\
\multicolumn{1}{|r}{Cluster}\vspace{-0.1cm}             & \multicolumn{3}{|c}{$a+b(\frac{R}{R_h})$         }                                    & \multicolumn{3}{c|}{$a+b\log(\frac{{\cal N}_*}{{\cal N}_{*h}})$}                           & \multicolumn{3}{c|}{$a+b\log(\frac{N_{{\rm dyn}}}{N_{{\rm dyn},h}})$}              & \multicolumn{3}{c|}{$N_{\rm dyn}=t_*/t_{\rm dyn}$}                     \\
\multicolumn{1}{|r}{} \vspace{-0.1cm}                   & \multicolumn{3}{|c}{}                                                                 & \multicolumn{3}{c|}{}                                                                      & \multicolumn{3}{c|}{}                                                              & \multicolumn{3}{c|}{}                     \\
\multicolumn{1}{|r}{}                                   & \multicolumn{1}{|c}{$p$-value} & \multicolumn{1}{c}{a}  & \multicolumn{1}{c|}{b}       & \multicolumn{1}{c}{$p$-value}   & \multicolumn{1}{c}{a}  & \multicolumn{1}{c|}{b}          & \multicolumn{1}{c}{$p$-value}   & \multicolumn{1}{c}{a}  & \multicolumn{1}{c|}{b}  & \multicolumn{1}{c}{$R_{60}$}  & \multicolumn{1}{c}{$R_{h}$} & \multicolumn{1}{c|}{$R_{max}$}                    \\[0.5ex]
\hline
Orion Nebula\vspace{-0.0cm}  &   \textcolor{gg}{0.009} & \textcolor{gg}{0.77} & \textcolor{gg}{ 0.19} &    \textcolor{gg}{0.007} &    \textcolor{gg}{1.02} &    \textcolor{gg}{-0.24} &  \textcolor{gg}{0.006} &    \textcolor{gg}{0.89} &    \textcolor{gg}{-0.32}  &   24.4  &  5.84  &  2.06  \\
Carina Nebula a)             &   \textcolor{gg}{0.002} & \textcolor{gg}{0.76} & \textcolor{gg}{ 0.22} &    \textcolor{gg}{0.002} &    \textcolor{gg}{1.02} &    \textcolor{gg}{-0.22} &  \textcolor{gg}{0.002} &    \textcolor{gg}{0.90} &    \textcolor{gg}{-0.29}  &   14.2  &  1.52  &  0.53  \\
Carina Nebula b)             &   \textcolor{gg}{0.003} & \textcolor{gg}{0.59} & \textcolor{gg}{ 0.39} &    \textcolor{gg}{0.002} &    \textcolor{gg}{1.00} &    \textcolor{gg}{-0.44} &  \textcolor{gg}{0.003} &    \textcolor{gg}{0.84} &    \textcolor{gg}{-0.57}  &   1.83  &  0.83  &  0.49  \\
Carina Nebula c)             &   \textcolor{rr}{0.509} & \textcolor{rr}{1.00} & \textcolor{rr}{-0.01} &    \textcolor{gg}{0.316} &    \textcolor{gg}{1.00} &    \textcolor{gg}{-0.06} &  \textcolor{gg}{0.309} &    \textcolor{gg}{0.98} &    \textcolor{gg}{-0.08}  &   1.95  &  0.51  &  0.26  \\
Eagle Nebula                 &   \textcolor{gg}{0.005} & \textcolor{gg}{0.73} & \textcolor{gg}{ 0.25} &    \textcolor{gg}{0.002} &    \textcolor{gg}{1.02} &    \textcolor{gg}{-0.31} &  \textcolor{gg}{0.002} &    \textcolor{gg}{0.86} &    \textcolor{gg}{-0.41}  &   3.83  &  1.08  &  0.42  \\
M 17                         &   \textcolor{gg}{0.051} & \textcolor{gg}{0.85} & \textcolor{gg}{ 0.13} &    \textcolor{gg}{0.003} &    \textcolor{gg}{1.01} &    \textcolor{gg}{-0.26} &  \textcolor{gg}{0.003} &    \textcolor{gg}{0.85} &    \textcolor{gg}{-0.35}  &   3.60  &  0.75  &  0.20  \\
NGC 6357 a)                  &   \textcolor{gg}{0.039} & \textcolor{gg}{0.55} & \textcolor{gg}{ 0.44} &    \textcolor{gg}{0.088} &    \textcolor{gg}{1.05} &    \textcolor{gg}{-0.34} &  \textcolor{gg}{0.092} &    \textcolor{gg}{0.94} &    \textcolor{gg}{-0.43}  &   3.35  &  0.57  &  0.25  \\
NGC 6357 b)                  &   \textcolor{gg}{0.273} & \textcolor{gg}{0.88} & \textcolor{gg}{ 0.09} &    \textcolor{gg}{0.214} &    \textcolor{gg}{0.98} &    \textcolor{gg}{-0.18} &  \textcolor{gg}{0.209} &    \textcolor{gg}{0.95} &    \textcolor{gg}{-0.23}  &   2.16  &  0.99  &  0.42  \\
NGC 6357 c)                  &   \textcolor{gg}{0.450} & \textcolor{gg}{0.98} & \textcolor{gg}{ 0.02} &    \textcolor{gg}{0.233} &    \textcolor{gg}{1.03} &    \textcolor{gg}{-0.13} &  \textcolor{gg}{0.213} &    \textcolor{gg}{0.98} &    \textcolor{gg}{-0.17}  &   2.96  &  0.40  &  0.23  \\
Lagoon Nebula                &   \textcolor{gg}{0.012} & \textcolor{gg}{0.46} & \textcolor{gg}{ 0.50} &    \textcolor{gg}{0.025} &    \textcolor{gg}{1.06} &    \textcolor{gg}{-0.45} &  \textcolor{gg}{0.024} &    \textcolor{gg}{0.93} &    \textcolor{gg}{-0.58}  &   6.70  &  1.15  &  0.59  \\
Rosette Nebula               &   \textcolor{gg}{0.100} & \textcolor{gg}{0.55} & \textcolor{gg}{ 0.47} &    \textcolor{gg}{0.116} &    \textcolor{gg}{1.03} &    \textcolor{gg}{-0.50} &  \textcolor{gg}{0.121} &    \textcolor{gg}{0.93} &    \textcolor{gg}{-0.60}  &   2.04  &  0.91  &  0.65  \\
W 3                          &   \textcolor{rr}{0.713} & \textcolor{rr}{1.10} & \textcolor{rr}{-0.08} &    \textcolor{rr}{0.806} &    \textcolor{rr}{0.99} &    \textcolor{rr}{ 0.14} &  \textcolor{rr}{0.814} &    \textcolor{rr}{1.05} &    \textcolor{rr}{ 0.18}  &   4.50  &  0.68  &  0.21  \\
NGC 6334                     &   \textcolor{gg}{0.050} & \textcolor{gg}{0.69} & \textcolor{gg}{ 0.26} &    \textcolor{gg}{0.089} &    \textcolor{gg}{1.01} &    \textcolor{gg}{-0.27} &  \textcolor{gg}{0.079} &    \textcolor{gg}{0.95} &    \textcolor{gg}{-0.35}  &   2.90  &  0.81  &  0.36  \\
NGC 1893                     &   \textcolor{gg}{0.001} & \textcolor{gg}{0.35} & \textcolor{gg}{ 0.62} &    \textcolor{gg}{0.001} &    \textcolor{gg}{1.03} &    \textcolor{gg}{-0.62} &  \textcolor{gg}{0.001} &    \textcolor{gg}{0.82} &    \textcolor{gg}{-0.79}  &   1.37  &  0.33  &  0.18  \\
NGC 3576                     &   \textcolor{rr}{0.933} & \textcolor{rr}{1.21} & \textcolor{rr}{-0.13} &    \textcolor{rr}{0.843} &    \textcolor{rr}{1.02} &    \textcolor{rr}{ 0.15} &  \textcolor{rr}{0.842} &    \textcolor{rr}{1.06} &    \textcolor{rr}{ 0.19}  &   2.47  &  0.87  &  0.19  \\
NGC 2264                     &   \textcolor{gg}{0.175} & \textcolor{gg}{0.65} & \textcolor{gg}{ 0.32} &    \textcolor{gg}{0.194} &    \textcolor{gg}{1.03} &    \textcolor{gg}{-0.37} &  \textcolor{gg}{0.192} &    \textcolor{gg}{0.94} &    \textcolor{gg}{-0.46}  &   2.34  &  0.99  &  0.65  \\
DR 21                        &   \textcolor{rr}{0.805} & \textcolor{rr}{1.12} & \textcolor{rr}{-0.11} &    \textcolor{rr}{0.708} &    \textcolor{rr}{1.00} &    \textcolor{rr}{ 0.07} &  \textcolor{rr}{0.696} &    \textcolor{rr}{1.01} &    \textcolor{rr}{ 0.08}  &   4.01  &  0.52  &  0.26  \\
RCW 38                       &   \textcolor{gg}{0.324} & \textcolor{gg}{0.94} & \textcolor{gg}{ 0.05} &    \textcolor{gg}{0.246} &    \textcolor{gg}{1.01} &    \textcolor{gg}{-0.09} &  \textcolor{gg}{0.240} &    \textcolor{gg}{0.97} &    \textcolor{gg}{-0.12}  &   9.63  &  1.48  &  0.38  \\
W 4                          &   \textcolor{gg}{0.132} & \textcolor{gg}{0.44} & \textcolor{gg}{ 0.57} &    \textcolor{gg}{0.165} &    \textcolor{gg}{1.06} &    \textcolor{gg}{-0.59} &  \textcolor{gg}{0.173} &    \textcolor{gg}{0.94} &    \textcolor{gg}{-0.72}  &   0.44  &  0.22  &  0.15  \\
NGC 2362                     &   \textcolor{gg}{0.040} & \textcolor{gg}{0.29} & \textcolor{gg}{ 0.58} &    \textcolor{gg}{0.037} &    \textcolor{gg}{0.96} &    \textcolor{gg}{-0.83} &  \textcolor{gg}{0.035} &    \textcolor{gg}{0.79} &    \textcolor{gg}{-1.07}  &   2.39  &  1.10  &  0.49  \\
Flame Nebula                 &   \textcolor{rr}{0.722} & \textcolor{rr}{1.19} & \textcolor{rr}{-0.16} &    \textcolor{rr}{0.755} &    \textcolor{rr}{1.00} &    \textcolor{rr}{ 0.26} &  \textcolor{rr}{0.755} &    \textcolor{rr}{1.06} &    \textcolor{rr}{ 0.35}  &   9.79  &  4.22  &  1.55  \\
RCW 36                       &   \textcolor{gg}{0.423} & \textcolor{gg}{0.99} & \textcolor{gg}{ 0.01} &    \textcolor{gg}{0.313} &    \textcolor{gg}{1.00} &    \textcolor{gg}{-0.04} &  \textcolor{gg}{0.310} &    \textcolor{gg}{0.99} &    \textcolor{gg}{-0.05}  &   7.38  &  1.90  &  0.54  \\
\hline
\end{tabular}
\tablecomments{$n_{\rm tot}$ is the total number of stars within $R_{\rm max}$, dividend in $N_{\rm a}$ annuli.
The Pearson's
correlation coefficient $p$-values for each cluster are assuming
the high resolution profile of $\delta_{\rm ADP,6}$ and circular
symmetry. Values $<0.5$ (green) indicate correlation in the expected direction. Cluster ages $t_*$ are from \citet{getman2014}. The dynamical ages $N_{\rm dyn}$ are lower limits (see text).}
\end{table*}

For each subcluster we refined the determination of the center using
an iterative method, as in \citetalias{dario2014}: we start from the
maximum of the nearest neighbor map, and recompute the center of
stellar positions recursively while decreasing the aperture size for
this computation until a convergence is found.

For a full description of the Angular Dispersion Parameter \adp\ we
refer the reader to \citetalias{dario2014}; here we summarize its
definition and characteristics. Given a circular (or elliptical)
aperture, or one of some concentric annuli enclosed in it, the area is
divided in $N$ sectors of equal area, each containing $n_i$
stars. \adp\ is then computed as:
\begin{equation}
\delta_{\rm ADP,N}=\sqrt{\frac{1}{(N - 1)\bar{n}}\sum_{i=1}^{N} (n_i - \bar{n})^2 }=\sqrt{\frac{\sigma^{2}}{\sigma^{2}_{Poisson}}}.
\end{equation}
\noindent This parameter is subject to a statistical error, which does
not depend only on the number of
sectors $N$, and can be decreased further by averaging all possible
orientations of the sector pattern \citepalias[see ][]{dario2014}. For
an isotropic centrally concentrated cluster, \adp$\simeq1$. Given a
substructured morphology the actual value of \adp$>1$
depends on the number of stars in each annular region, which therefore
must be fixed for a meaningful comparison between regions. We choose 60 stars per annulus as a compromise
between statistical significance of number counts, and radial
resolution. An example of the subdivision of the ONC population in
circular or elliptical annuli is shown in the first panel of Figure
1. Depending on $N$, \adp\ measures a given
angular mode of substructure. As default for our analysis we consider
$N=6$, but also test the effect in our results assuming
$N=4$ or 9.

As shown in \citetalias{dario2014}, ellipticity causes \adp\ to
increase even for a smooth distribution, mimicking the effect of
substructure, and, e.g., was found to be a
main contributor to the measured \adp$>1$. This is because at a given
distance (in circular symmetry) from the center the number counts are
higher along the major axis and lower along the minor axis. A similar
bias was also shown to affect the MST ${\cal Q}$ parameter
\citep{bastian2009}. Thus in \citetalias{dario2014} we generalized the
\adp\ method to measure subclustering adopting elliptical symmetry --
i.e., dividing the population into elliptical concentric sectors.
For the sample here, we thus determined elongation
by fitting simple 2D Plummer
models to each cluster, leaving ellipticity and semi-major axis
position angles as free parameters. In Figures \ref{figure:radec1} and
\ref{figure:radec2} we display the 18 MYStiX regions, highlighting the
subclusters selected for this study. The red circles denote the
maximum useful apertures ($R_{\rm max}$), and the blue ellipse the
best fit elliptical model inside each of them.

We measure the radial profile of $\delta_{\rm ADP,6}$ for our
identified MYStIX subclusters. Given the intrinsic difference in their
number of members (from $\sim300$ to $\sim 1800$ members, see Table
\ref{table}), we adopt two choices: a) the \emph{high resolution}
profile, in which for both circular and elliptical symmetry we divide
the populations in all concentric annuli containing 60 stars within
$R_{\rm max}$, and b) the {\em low resolution} profile, where we
consider 3 radii: 1) $R_{60}$, i.e., the smallest circle from each
center containing 60 stars; 2) the mean of $\delta_{\rm ADP,6}$ within
the half-mass radius $R_h$; 3) the mean of $\delta_{\rm ADP,6}$ within
the cluster radius $R_{\rm max}$. Since the MYStIX fields of view are
not uniform, and because of our constraints (\S\ref{section:data}) to
avoid overlapping of subclusters in regions with multiple central
concentrations, both $R_{\rm max}$ and $R_h$ retain some
arbitrariness. Also, the MYStIX cluster sample is heterogenous in that
individual regions, depending on their distance and dust extinction,
have different levels of (in)completeness. As we show in
\S\ref{section:results}, such limitations are insufficient to erase
the general qualitative trends we detect.

\section{Results}
\label{section:results}

In Table \ref{table} we report for each cluster the three radial bins of our
low resolution profile, together with the total number of stars,
annuli, and the average $\langle\delta_{\rm ADP,6}\rangle$ within $R_{\rm
  max}$, all assuming circular symmetry. In nearly all cases
$\delta_{\rm ADP,6}>1$, indicating that none of these regions has reached
an isotropic spatial distribution of sources. In some cases (e.g., the
clustered concentration of members in the Rosetta nebula), the value
is fairly small; in others (e.g., NGC1893) $\delta_{\rm ADP,6}$ traces the
visibly irregular spatial distribution of sources.

We look for systematic correlations in the degree of substructure with
distance from the center and with surface density, employing the
Pearson product-moment correlation coefficient and the Spearman
correlation rank statistics. Specifically, we would expect that the
core of the clusters, where the density is higher, would have reached
a smoother distribution due to a higher rate of stellar interactions
(scaling as the dynamical time or the free-fall time, $t_{\rm ff}\propto
\rho^{-1/2}$). In Table \ref{table} we also report
the one-sided $p-$values corresponding to a chance probability that
either a positive trend of $\delta_{\rm ADP,6}$ versus $R$, or a negative
trend of $\delta_{\rm ADP,6}$ versus the surface density of stars per
pc$^2$, ${\cal N}_*$, adopting the high-resolution $\delta_{\rm ADP}$
profile. Indeed, for the majority of the clusters we identify a
correlation, in some cases very significant ($p<1\%$).

We also analyze these correlations considering all regions at once. In
Figures \ref{figure:adp_vs_radii} and \ref{figure:adp_vs_density} we
present the dependence of $\delta_{\rm ADP,6}$ respectively on $R$ and
$\log{\cal N}_*$, using both our low- and high-resolution profiles, for both the assumptions of circular
and elliptical apertures. As indicated by the $p-$values, the
correlations become more significant, due to the
increased number statistics, in particular in the high resolution
$\delta_{\rm ADP}$ profiles. These trends become slightly weaker when
elliptical symmetry is adopted; this is because, as found in the ONC
in \citetalias{dario2014}, this assumption naturally removes the first
order of anisotropy in the stellar distribution, i.e., the presence of
an elongation in a particular direction.

In Table \ref{table} we also show for comparison the MST ${\cal Q}$
parameter, computed for each cluster within $R_{h}$ and
$R_{\rm max}$. Unlike for \adp, where significant variations are measured
within each cluster and among different clusters, all ${\cal Q}$ values
cluster around 0.8, the value generally assumed to discriminate
between clumpy (${\cal Q}<0.8$) and centrally concentrated ${\cal
  Q}>0.8$) stellar distributions \citep{cartwright2004}. This confirms
that the ${\cal Q}$ parameter is adequate to separate highly
substructured regions, where individual clumps are evident, but has
little or no diagnostic power
for probing more moderate degrees of substructure within
centrally-concentrated clusters,
where \adp\ is far more sensitive
\footnote{For a smooth isotropic stellar distribution within a sphere, $\cal{Q}$
increases with central concentration: ${\cal Q}\simeq0.8$ for constant
volume density $n\propto r^0$ within a sphere; ${\cal Q}\simeq0.95$
for $n\propto r^{-2}$}.  Also, interestingly, in most of our clusters
${\cal Q}$ tends to be slightly larger at $R_{\rm max}$ than at
$R_{h}$, indicating an opposite trend at that we detect from \adp.

Considering the heterogeneity of the regions included in this
analysis, both in terms of completeness (\S\ref{section:analysis}) and
because each region may differ in their physical and dynamical age, we
also test these correlations normalizing each quantity on some
characteristic value typical of each cluster. Specifically, we
consider $\delta_{\rm ADP}^\prime$, which is $\delta_{\rm ADP}$
normalized on its average for each particular cluster; also, the
radius $R^\prime$ is $R/R_h$ and the surface density ${\cal
  N}_*^\prime={\cal N}_*(R)/{\cal N}_*(R_h)$. The result is shown on
the right-hand panels of Figures \ref{figure:adp_vs_radii} and
\ref{figure:adp_vs_density}. Such normalizations always tighten the
correlations, especially in the low-resolution case. We also tested if
this improvement is particularly due to the normalization on one
specific axis, considering all the combinations of such
normalization. We found that for the $\delta_{\rm
  ADP}^\prime(R^\prime)$ trend, the normalization on $\delta_{\rm
  ADP}^\prime$ is the dominant factor, indicating that the MYStIX
clusters may have inherent differences in their overall degree of
substructure.
On the other hand, both normalizations in $\delta_{\rm
  ADP}^\prime(\log {\cal N}_*^\prime)$ contribute to the significance
of the correlation between these quantities, probably due to the
effect of inconsistent completeness affecting the densities we
measure.  From the linear fits of the high-density profiles, assuming
circular symmetry, we measure: $\delta_{\rm ADP,6}^\prime=0.84+0.14
R/R_h$ and $\delta_{\rm ADP,6}^\prime=1.01-0.19\log {\cal N}_*/{\cal
  N}_{*h}$, which are quantitative relations that can be compared to
simulations of the dynamical evolution of young clusters.

Last, changing the numbers of sectors to 4 or 9 does not affect these
trends. Solely, as observed by \citetalias{dario2014},
increasing the number of sectors tends to decrease $\delta_{\rm ADP}$,
indicating that the substructure of young stellar clusters is
dominated by anisotropy on larger angular scales.

We also looked for trends of \adp\ with the median cluster age from
\citet{getman2014b}, derived from the comparison of X-ray-derived
stellar masses and extinction corrected $J-$band magnitudes for a
sample of sources in each region. We detected no significant
correlation for any choice of \adp\ for each cluster. This may be due
to the fact that the cluster ages differ by up to a factor of 3 -
compared to a much larger range of $dynamical$ ages as a function of
distance from the center within each cluster - and are in general very
uncertain \citep[e.g.,][]{dario2010,soderblom2013}.

These results are amongst the first to probe dynamical evolution
during and shortly after star cluster formation, specifically the
dissolution of substructure in embedded systems with membership based
on X-ray emission. Compared to other observational studies
\citep{gutermuth2005,gutermuth2008,ascenso2007,schmeja2008,getman2014,beccari2014},
which have been able to qualitatively detect systematic differences in
the location of younger and older stars in young clusters, and their
morphology as a function of age, our results illustrate how \adp\ can
be a metric of relative local dynamical age, which is greater in
central, denser regions.

For gravitationally bound, virialized systems
with virial parameter $\alpha_{\rm vir}\equiv 5 \sigma^2 R/(GM)$,
where $\sigma$ is the 1D mass-averaged velocity dispersion, we can
assess the dynamical age $t_*/t_{\rm dyn}$. Here $t_*$ is the actual
age of the system and $t_{\rm dyn}\equiv R/\sigma$ is the local dynamical
time. Thus
\begin{equation}
\frac{t_*}{t_{\rm dyn}} = 1.06 \alpha_{\rm vir}^{1/2} \left(\frac{t_*}{2\:{\rm Myr}}\right)
\left(\frac{\Sigma}{100\:M_\odot\:{\rm pc^{-2}}}\right)^{1/2} \left(\frac{R}{{\rm pc}}\right)^{-1/2},
\end{equation}
\noindent where $\Sigma\equiv M/(\pi R^2)$. The conversion between ${\cal N}_*$
and $\Sigma$ depends on the mean stellar mass of the X-ray-detected
sources, the degree of incompleteness of the stellar population and
the contributions of gas. For a near complete stellar census and a
similar mass of gas and stars we expect $\Sigma \simeq ({\cal
  N}_*/{\rm pc}^{-2}) M_\odot\:{\rm pc}^{-2}$. With the exception of
the ONC, most of the MYStIX clusters are expected to be more
incomplete in their stellar census, due to their farther distances, and lower Chandra exposure times in X-rays.
As described in \citet{feigelson2013}, however, the characterization of the completeness in each MYStIX region is challenging because the membership assignment from \citet{broos2013} is quite restrictive and the completeness depends also on IR data and the reddening and crowding on each region.

Still, using this conversion factor we may derive a lower limit on the
dynamical age at $R_{\rm 60}$, $R_h$ and $R_{\rm max}$. We have
adopted the median age for each region from \citet{getman2014}, which
is available for all except four of our subclusters, which have been
assigned a typical age of 2~Myr. We have also adopted $\alpha_{\rm
  vir}=1$ as a fiducial value.

Values $N_{\rm dyn}=t_*/t_{\rm dyn}\gtrsim 1$ then indicate this
region of the cluster is older than a crossing time, and thus likely
to be gravitationally bound and subject to dynamical smoothing of
substructure. As reported in Table \ref{table}, all our clusters
appear older than 1~$t_{\rm dyn}$ at least in the core region
($R_{60}$), and in some cases up to $R_h$. This conclusion is
reinforced considering that our estimates of $N_{\rm dyn}$ are lower
limits due to incompleteness.

Figure \ref{figure:adp_vs_dynage} shows the profiles of \adp\ as a
function of $N_{\rm dyn}$; we find that $\delta_{\rm adp}$ decreases
with increasing dynamical age, and reaches values close to unity
after a few dynamical times (although values of $N_{\rm dyn}$ here are
likely lower limits).
Despite the uncertainties in $\alpha_{\rm vir}$ and
$\Sigma$ for each cluster, the correlation p-values in Figure
\ref{figure:adp_vs_dynage} are always smaller than those derived for
\adp\ as a function of $R$ and ${\cal N}_*$, indicating a tighter fit.
As for the other parameters, Table \ref{table} also reports the best
fit simple function
correlation results for each individual cluster, showing a
behavior similar to that of \adp\ versus ${\cal N}_*$.

Measuring the degree of substructure was proposed as a way to assess
whether star cluster formation was dynamically slow or rapid
\citep{tan2006}. We expect the quantitative dependence of
\adp$(R,{\cal N}_*,t/t_{\rm dyn})$ is an important constraint against which to test
theoretical models of star cluster formation.

As we have shown, \adp\ constitutes a much more sensitive estimator of
substructure within centrally concentrated clusters or subclusters
than the MST ${\cal Q}$ parameter, and advocate its use in both
observational and theoretical
studies of early cluster evolution.

The absolute ages and age spreads of young star clusters are very
difficult to measure and current estimates contain significant
uncertainties. Still, the results of future studies assessing such
ages and age spreads of these star clusters systematically can be
compared with our \adp$(R,{\cal N}_*,t/t_{\rm dyn})$ results to place the discovered
structural evolution on an absolute timescale.

\acknowledgements
KOJ and JCT acknowledge support from NASA Astrophysics Theory and Fundamental Physics grant ATP09-0094. NDR acknowledges support from the Theory Postdoctoral Fellowship from the University of Florida Department of Astronomy and College of Liberal Arts and Sciences.

\end{document}